\documentclass[a4paper,11pt]{article}
\pdfoutput=1 
\usepackage{jheppub} 
\usepackage[utf8]{inputenc}
\usepackage{amsthm}
\usepackage{mathrsfs} 
\usepackage[all]{xy}
\graphicspath{{./images/}}
\theoremstyle{plain}

\theoremstyle{definition}

\numberwithin{thm}{section}

\newcommand{\vev}[1]{ \left\langle {#1} \right\rangle }
\newcommand{\bra}[1]{ \langle {#1} | }
\newcommand{\ket}[1]{ | {#1} \rangle }

\def\Im{\mathop{\mathrm{Im}}}
\usepackage{slashed}

\def\CA{{\cal A}}

\def\CH{{\cal H}}

\def\CM{{\cal M}}

\def\CO{{\cal O}}

\def\BR{{\mathbb R}}

\def\beq#1\eeq{\begin{align}#1\end{align}}

\title{ Black hole information and
Reeh-Schlieder theorem
}

\preprint{}

\author{Kazuya Yonekura}
\affiliation{Faculty of Arts and Science, Kyushu University, 
 Fukuoka, Fukuoka, 819-0395, Japan
}

\abstract{
The Reeh-Schlieder theorem, with the time-slice axiom of quantum field theory, may be used
to recover the information which falls into a black hole. Analyticity of quantum fields in states with finite energy plays the crucial role.
In AdS spacetime, our argument based on the Reeh-Schlieder theorem is consistent with the argument that there is no information loss because of the AdS/CFT correspondence.
} 

\begin{document}

\maketitle
\section{Introduction}
The black hole information problem~\cite{Hawking:1976ra} is roughly described as follows.
We throw something into a black hole. Eventually the black hole evaporates. Naively it seems that
the process of Hawking radiation~\cite{Hawking:1974sw} is independent of what we throw into the black hole except for
its mass, charge, and angular momentum. Therefore the information is lost. 

However, what do we mean by ``information which is thrown into a black hole"? One may consider that it is a state vector associated to the matter which is thrown into the black hole.
This is partially quantum mechanical description of information.

In complete quantum mechanical description in which the position of the matter is also described quantum mechanically, and in particular in
relativistic quantum field theory, the situation is much more subtle.
It is not easy to see how to make precise the concept of ``information localized in some region". 
(See Appendix~\ref{app:wave} for a simple quantum mechanical discussion.)
Indeed, the theorem by Reeh and Schlieder~\cite{RS} (see \cite{Streater:1989vi,Witten:2018zxz} for reviews) may be interpreted as a fact that ``information" cannot be
localized precisely. Then we lose a clear distinction between ``information inside the black hole" and ``information outside the black hole".
This idea has essentially appeared as a construction of the black hole interior 
by using the degrees of freedom outside of it~\cite{Papadodimas:2013wnh,Papadodimas:2013jku,Verlinde:2013qya}.

The purpose of this paper is to try to make the  preservation of information in the black hole evaporation more precise in this context.
From the beginning we assume the existence of a state in which the time evolution is unitary.
Then, we add perturbation to this situation by introducing additional matter which is thrown into the black hole, and formulate how that information is recovered 
after the evaporation of the black hole. Therefore, we are going to argue that there is no information loss in the perturbation of a given state which is assumed to be unitary.
A big black hole may be constructed by adding perturbation to a small black hole by throwing a small amount of matter step by step, 
so if our proposal is correct, the information problem may be reduced to that of a Planck scale black hole, which requires a full theory of quantum gravity.
As we will see, our argument generalizes the fact that there is no information loss in black holes in AdS spacetime because there is no information loss in the dual CFT.

\section{Reeh-Schlieder theorem }\label{sec:RS}
In this section we review the Reeh-Schlieder theorem.
In flat Minkowski space, a rigorous proof and an excellent review are already avaliable in \cite{Streater:1989vi} and \cite{Witten:2018zxz}.
In curved backgrounds, we only give sketchy discussions without trying to give a complete justification (see \cite{Strohmaier:2002mm,Gerard:2017apb,Sanders:2008gs, Morrison:2014jha} for some developments).

\subsection{Flat Minkowski space}
Let $\phi(x)$ represent local operators of the theory (or more precisely operator valued tempered distribution),
let $\phi(f)=\int d^d x f(x) \phi(x)$ represent smeared operators for (Schwartz) test functions $f(x)$ on $\BR^{1,d-1}$,
and let $\CA(\CO)$ be the algebra of operators generated by $\phi(f)$ for test functions $f(x)$ which have support in 
the open set $\CO \subset \BR^{1,d-1}$. 
We may more simply say that $\CA(\CO)$ is the set of all operators in the region $\CO \subset \BR^{1,d-1}$ of the Minkowski spacetime $\BR^{1,d-1}$.

Let $\widetilde{\CH}$ be a subspace of the total Hilbert space $\CH$ with the following property.
Consider operators $\exp ( \epsilon H )$ where $H$ is the Hamiltonian and $\epsilon >0$ is a positive real number.
Then, a state vector $\ket{\Psi} $ is defined to be an element of $\widetilde{\CH}$ if $\ket{\Psi}$ is in the domain of definition
of the operator $\exp ( \epsilon H )$ for some $\epsilon$. This basically means that $\exp ( \epsilon H )\ket{\Psi}$ has a finite norm
$\bra{\Psi} \exp(2\epsilon H) \ket{\Psi} < + \infty$. We call this condition as the finite energy condition. 
A sufficient condition is that the state $\ket{\Psi}$ only contains eigenstates of $H$ bounded by some upper bound $E_0$.
However, we remark that this is not a necessary condition. For example, states which look like a thermal state, e.g. $\ket{\Psi} \sim \sum_n e^{-\beta E_n/2} \ket{n}$,
satisfy the condition by taking $\epsilon$ to be such that $2\epsilon < \beta$.

Let us fix a state $\ket{\Psi} \in \widetilde{\CH}$.
The Reeh-Schlieder theorem states that the space of states
\beq
\{ a \ket{\Psi}  ;~ a \in \CA(\CO) \}
\eeq
is dense in the Hilbert space $\CH$. Namely, any state in the Hilbert space is well approximated by a state of the form $a \ket{\Psi}$ to an arbitrary good accuracy.

A sketch of the proof goes as follows. Suppose that the statement of the Reeh-Schlieder theorem does not hold.
Then there exists a state $\ket{\chi}$ which is orthogonal to all of $a \ket{\Psi}$,
\beq
\bra{\chi}\phi(x_1) \phi(x_2) \cdots \phi(x_n) \ket{\Psi}=0, \qquad x_i \in \CO~(i=1,\cdots,n).
\eeq
Now, notice that the above matrix elements can be written as
\beq
& \bra{\chi}\phi(x_1) \phi(x_2) \cdots \phi(x_n) \ket{\Psi} \nonumber \\
=&  \bra{\chi} e^{- i x_1  P }\phi(0) e^{ i (x_1-x_2)  P } \phi(0) \cdots  e^{ i (x_{n-1}-x_n) P} \phi(0) e^{ i x_n P} \ket{\Psi} .
\eeq
where $P=P^\mu=(H,\vec{P})$ is the four-momentum operator, and $xP$ is the inner product $x P=x_\mu P^\mu= -x^0 H+\vec{x}\cdot \vec{P}$.
We analytically continue $x_i$ as
$
x_i \to z_i=x_i - iy_i
$
in such a way that vectors
$-y_1 $ and $ (y_i - y_{i+1})$
are inside the forward light cone. Notice that in the analytic continuation
we have factors $e^{- y_1  P }$, $e^{(y_i-y_{i+1})P}$, and $e^{y_n P}$ in the above equation. The factors
$e^{- y_1  P }$ and $e^{(y_i-y_{i+1})P}$ are bounded operators with exponential damping of high energy modes.
Therefore, the analytic continuation is safe as far as these factors are concerned.
The only factor which is dangerous is $e^{ y_n P}$, which is unbounded and exponentially large for high energy modes because $-y_n$ is in the forward light cone.
However, the spectral condition $H \geq |\vec{P}|$ implies that $y_n P \leq (|y_n^0| +|\vec{y}_n| )H$.
By definition of $\ket{\Psi}$, there exists $\epsilon$ such that $e^{\epsilon H}\ket{\Psi}$ is finite. Therefore, as long as $y_n$ satisfies
 $(|y_n^0| +|\vec{y}_n|) < \epsilon$,
we can make sense of the state vectors 
$
e^{ y_n P}\ket{\Psi} = e^{ y_n P - \epsilon H} ( e^{\epsilon H} \ket{\Psi} )
$
because $e^{ y_n P - \epsilon H} $ is bounded.
Hence we can safely perform the analytic continuation in this region.

Thus we see that 
\beq
f(z_1, \cdots, z_n) =  \bra{\chi}\phi(z_1) \phi(z_2) \cdots \phi(z_n) \ket{\Psi} \label{eq:holo}
\eeq
is holomorphic in the region where $\Im z_1=-y_1$ and $-\Im (z_i-z_{i+1}) =y_i-y_{i+1}$ are inside the forward light cone and  $(|y_n^0| +|\vec{y}_n|) < \epsilon$.
Moreover, it is zero when ${\Im z_i}=0$ and $x_i \in \CO$. Roughly speaking, such a function is analytically continued to be zero everywhere; 
see \cite{Streater:1989vi,Witten:2018zxz} for more rigorous treatment. By the analytic continuation,
we conclude that $f(x_1, \cdots,x_n)$ is zero not only when $x_i \in \CO$, but for any $x_i $ in the Minkowski space.

\begin{figure}
\centering
\includegraphics[width=.35\textwidth]{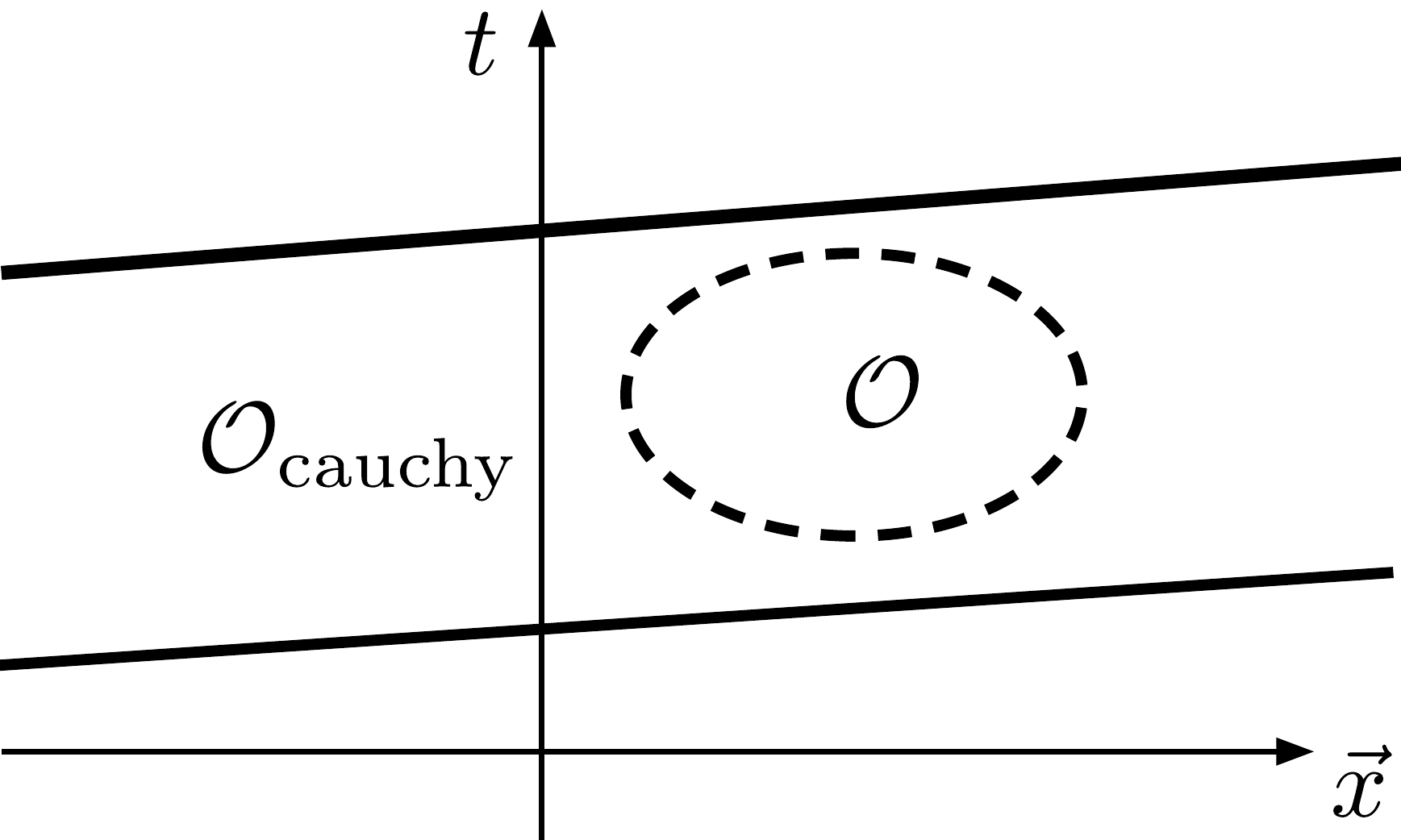}
\caption{Extending $\CO$ to $\CO_{\rm cauchy}$ \label{fig:cauchy}}
\end{figure}

In particular, let us take an open set $\CO_{\rm cauchy}$ which is a tubular neighborhood of a Cauchy surface of the spacetime
such that it contains $\CO$; see Figure~\ref{fig:cauchy}. 
An example is given by $\CO_{\rm cauchy}=\{ (t, \vec{x});~ t_1 < t <t_2 \}$.
Now we have the following axioms of quantum field theory. 
(See the axioms F. Completeness and G. ``Time-slice Axiom" of Section II.1.2 of \cite{Haag:1992hx}. See also
Theorem~4-5 of \cite{Streater:1989vi}.)
The axioms say that the algebra $\CA(\CO_{\rm cauchy})$ acts irreducibly on the Hilbert space $\CH$ and hence we can generate a dense set
in the Hilbert space by acting $\CA(\CO_{\rm cauchy})$ to $\ket{\Psi}$.\footnote{When there are conserved gauge charges such as the electric charges of QED,
the statement is expected to hold for each superselection sector in which the total charge is fixed. In the following, we always work in such
a superselection sector.}
By the above analytic continuation,
$\ket{\chi}$ is still orthogonal to this dense space and hence $\ket{\chi}=0$. This completes the sketch of the proof.

In the above sketch of the proof, the fact that the finite energy condition implies the analyticity was the important ingredient.
This point can be demonstrated by more simple quantum mechanical wave functions. See Appendix~\ref{app:wave} for the discussions.

There is an interesting corollary of the Reeh-Schlieder theorem.
Let $\CO_A$ be an open set of the spacetime and suppose there exists another open set $\CO_B$ which is space-like separated from $\CO_A$.
Then we have $a\ket{\Psi} \neq 0$ for any $a \neq 0$. The proof is as follows. Suppose $a\ket{\Psi}=0$. Then for any $b \in \CA(\CO_B)$
we get $ab\ket{\Psi}=\pm ba\ket{\Psi} = 0$. The sign $ \pm$ depends on whether the operators are bosonic/fermionic.
(Here we have assumed that $\ket{\Psi}$ has a definite fermion parity and hence without loss of generality we can assume that $a$ and $b$
also has definite fermion parities.) Now $b\ket{\Psi}$ spans a dense subset of the Hilbert space, and hence $ab\ket{\Psi}=0$ for any $b$
implies that $a=0$. This completes the proof.

\subsection{Remark on intuitive understanding and its surprising failure}

One intuitive way to understand the Reeh-Schlieder theorem may be as follows, and it is used in the standard understanding of the properties of the Rindler space.
Let $A$ and $B$ be spatial regions without overlap at a fixed time. We denote the causal diamonds of them in the spacetime as $\CO_A$ and $\CO_B$.
We may try to associate Hilbert spaces $\CH_A$ and $\CH_B$ to the regions $A$ and $B$, and the Hilbert space of the region $A \sqcup B$
may be thought to be factorized as $\CH_{A \sqcup B} =\CH_A \otimes \CH_B$ which follows from the intuition from lattice regularization (at least in the absence of gauge fields). 
We further suppose that these Hilbert spaces are all finite dimensional, again from the intuition from lattice in which each site has qubits.

Let us take a state $\ket{\Psi_{A \sqcup B}} \in \CH_{A \sqcup B}$. The Reeh-Schlieder theorem may be modeled in this finite dimensional setting
by saying that for any state $\ket{\Phi_{A \sqcup B}} $,
we can find a unique operator $a \in \CA(\CO_A)$ which acts on the Hilbert space $\CH_A$ such that $\ket{\Phi_{A \sqcup B}} = a \ket{\Psi_{A \sqcup B}}$.
Here in the right hand side, $a$ is an abbreviation of $a \otimes 1_B$ where $1_B$ is the identity operater acting on $\CH_B$.

We can also slightly rephase it as follows. For any $b \in \CA(\CO_B)$ which acts on $\CH_B$, we define a state as $\ket{\Phi_{A \sqcup B}} =b \ket{\Psi_{A \sqcup B}} $.
Then we can find $a \in \CA(\CO_A)$ such that $b \ket{\Psi_{A \sqcup B}} = a \ket{\Psi_{A \sqcup B}}$. The role of $A$ and $B$ can be interchanged.
This is possible, in the finite dimensional case, if and only if $\dim \CH_A = \dim \CH_B$ and that
$\ket{\Psi_{A \sqcup B}}$ is given as
\beq
\ket{\Psi_{A \sqcup B}} = \sum_{ij} c_{ij}\ket{\psi_A^i} \otimes \ket{\psi_B^j}
\eeq
where $\ket{\psi_A^i} $ and $ \ket{\psi_B^j}$ are basis vectors of $\CH_A$ and $\CH_B$ respectively, and the coefficients $c_{ij}$
have the maximal rank as a matrix $(c_{ij})$. In such a case, we may say that
the state $\ket{\Psi_{A \sqcup B}} $ is fully entangled, following the terminology of \cite{Witten:2018zxz}.

\begin{figure}
\centering
\includegraphics[width=.8\textwidth]{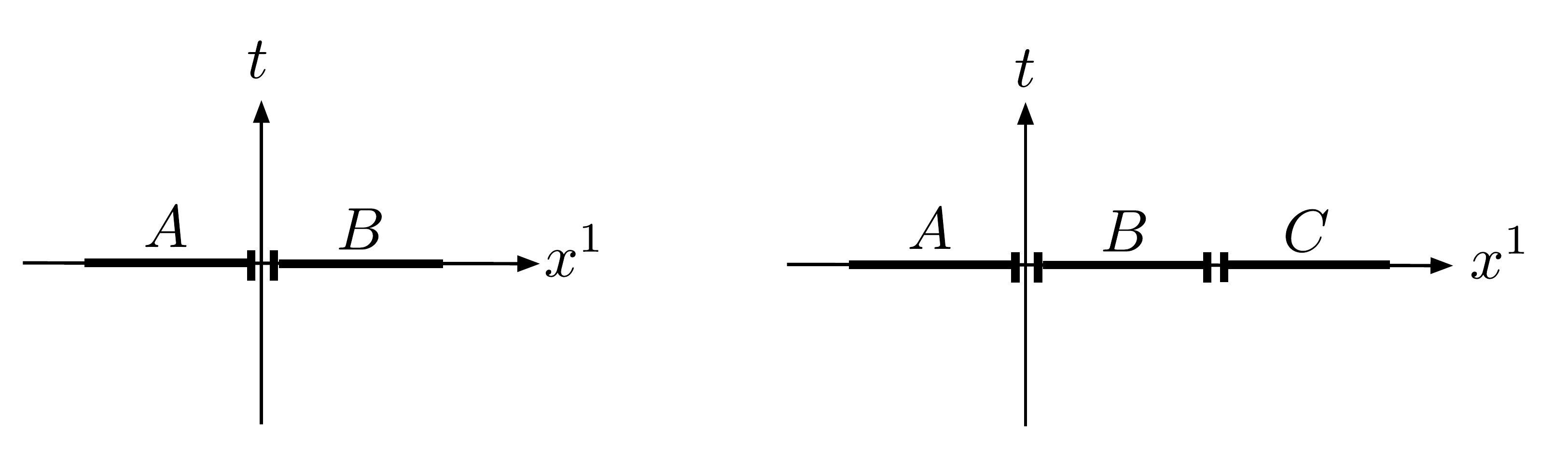}
\caption{Left: The decomposition of the space into regions $A$ and $B$. The causal diamond of $B$ gives a Rindler space.
Right: Dividing the space into three regions.  \label{fig:rindler}}
\end{figure}

The above understanding is standard in the Rindler space. In this case, we take $A=\{\vec{x}; ~ x^1<0 \}$ and
$B=\{\vec{x};~x^1>0 \}$. See the left side of Figure~\ref{fig:rindler}.
The vacuum state $\ket{\Omega}$ satisfies the condition $\ket{\Omega} \in \widetilde{\CH}$ of the Reeh-Schlieder theorem,
and hence, intuitively, the regions $A$ and $B$ are fully entangled. More explicitly, in the picture of the tensor factorization of the Hilbert space $\CH_A \otimes \CH_B$,
the vacuum is schematically given as $\ket{\Omega} \propto \sum_i e^{- \pi \omega_i} \ket{i}_A \otimes \ket{i}_B$, where $\ket{i}_A$ and $\ket{i}_B$ are eigenmodes of the boost operator $K$,
and $\omega$ is the (negative of) the eigenvalue of $K$ on $\ket{i}_B$ ($\ket{i}_A$).
See e.g. \cite{Harlow:2014yka} for a review.
This is the origin of the thermal properties of the Rindler space after tracing out the $\CH_A$.

However, the above intuition should not be trusted too much, as is well known to experts (see e.g. \cite{Witten:2018zxz} and references therein).
We give a rather surprising demonstration of its failure by using the right side of Figure~\ref{fig:rindler}. 
We divide the space into three regions
\beq
A=\{\vec{x}; ~ x^1<0 \}, \qquad B=\{\vec{x}; ~ 0 < x^1< \ell\}, \qquad C=\{\vec{x}; ~ x^1> \ell \}
\eeq
for some $\ell>0$. Consider a state $\ket{\Psi_{A \sqcup B \sqcup C}}$ for which the finite dimensional version of the Reeh-Schlieder theorem is supposed to hold.
Then, for any $b \in \CA(\CO_B)$, we can find an operator $a \in \CA(\CO_A)$ such that
\beq
a\ket{\Psi_{A \sqcup B \sqcup C}} = b \ket{\Psi_{A \sqcup B \sqcup C}}.
\eeq
Conversely, for any $a \in \CA(\CO_A)$, we can find an operator $b \in \CA(\CO_B)$ satisfying the above equation.
By a small computation and Shur's lemma, one can see that this is possible for the tensor-factorized finite dimensional Hilbert space only if
\beq
\ket{\Psi_{A \sqcup B \sqcup C}} = \left( \sum_{ij} c_{ij}\ket{\psi_A^i} \otimes \ket{\psi_B^j} \right) \otimes \ket{\Psi_C},
\eeq
where $c_{ij}$ has the maximal rank and $\ket{\Psi_C}$ is a state vector of $\CH_C$. 
Namely, $A$ and $B$ are fully entangled, and there is no entanglement between $A \sqcup B$ and $C$.
However, we can repeat the same argument for $B$ and $C$ by using 
\beq
b\ket{\Psi_{A \sqcup B \sqcup C}} = c \ket{\Psi_{A \sqcup B \sqcup C}}.
\eeq
and also for $A$ and $C$ by using
\beq
c\ket{\Psi_{A \sqcup B \sqcup C}} = a \ket{\Psi_{A \sqcup B \sqcup C}}.
\eeq
Then we conclude that $A$ is fully entangled with $B$, $B$ is fully entangled with $C$, and $C$ is fully entangled with $A$, which is impossible!
This clearly shows that the oversimplification of the Hilbert space $\CH$ by the tensor factorization in terms of finite dimensional Hilbert spaces $\CH_A,\CH_B, \CH_C$ can fail in the context of
information theory.

It is quite tempting to think about the implications of the above discussion for the ``entanglement monogamy problem" of the firewall paradox~\cite{Almheiri:2012rt}
and some of its proposed resolutions 
(e.g.~\cite{Papadodimas:2012aq, Maldacena:2013xja, Bousso:2012as, Harlow:2013tf, Papadodimas:2013wnh,Papadodimas:2013jku,Verlinde:2013qya}). 
See \cite{Harlow:2014yka} for a review.
The Reeh-Schlieder theorem was discussed in this context in \cite{Papadodimas:2013jku} (see also \cite{Chakraborty:2017pmn}).
We do not investigate this problem further to avoid dangerous statements,\footnote{It is tempting to formally call
$\CH_A$ as ``the degrees of freedom inside the event horizon", $\CH_B$ as ``the degrees of freedom in the late Hawking radiation", and $\CH_C$ as
``the degrees of freedom in the early Hawking radiation", and claim that they are fully entangled with each other by the Reeh-Schlieder theorem, 
despite the fact that it is impossible in the tensor-factorized finite dimensional Hilbert space. Of course these formal definitions are not exactly what is meant in the firewall paradox, but
it suggests that a careful definition of various concepts can be important. }
but the above discussion suggests a possibility that ``relativistic quantum information theory" rather than just an ordinary non-relativistic quantum information theory might
play a role in black hole information problems.

\subsection{Curved backgrounds} \label{sec:Curved}
We do not try to give a proof of the Reeh-Schlieder theorem on curved backgrounds (see \cite{Strohmaier:2002mm,Gerard:2017apb,Sanders:2008gs, Morrison:2014jha} for results in this direction).
However, we would like to make a few comments.

One of the essential points in the proof of the Reeh-Schlieder theorem was the analytic continuation of matrix elements like \eqref{eq:holo}.
This analytic continuation was made possible by the following two facts; (i) The energy of any system is bounded from below so that $e^{-\tau H}$ is a bounded operator for 
any positive $\tau$;
(ii) We choose state vectors $\ket{\Psi} \in \widetilde{\CH}$ such that $e^{\epsilon H} \Psi$ has a finite norm for some positive $\epsilon$.

On curved manifolds, there are no translation symmetries and hence the above proof does not apply straightforwardly.
However, even in curved backgrounds, we still expect that the ``energy" (in some appropriate sense) has a lower bound.

How about the condition that there exists $\epsilon >0$ such that $e^{\epsilon H} \ket{\Psi}$ has a finite norm?
To get some intuition about this problem, let us ask the following simpler question. We treat the metric as background fields.
Suppose that the background metric has a flat space form in the limit $t \to \pm \infty$. 
In the intermediate spacetime region, the metric deviates from the flat space. Now, we can prepare a state $\ket{\Omega_{t=-\infty}}$ in the Heisenberg picture
such that it is indeed in the vacuum state in the limit $t \to -\infty$.
This does not conicide with the state $\ket{\Omega_{t=+\infty}}$ which goes to the vauum in the limit $t \to + \infty$.
In free field theory, these two are related by using the Bogoliubov transformation of creation and annihilation operators
as in the Hawking's computation of the black hole radiation~\cite{Hawking:1974sw}. 
Now we ask the question of whether $\ket{\Omega_{t=-\infty}}$
satisfies the condition of the Reeh-Schlieder theorem in the region $t \to +\infty$
with respect to the Hamiltonian $H_{t =+\infty}$.

To make the question much simpler, let us investigate the above question in the case of the harmonic oscillator with a time-dependent Hamiltonian
\beq
H(t)=\frac{1}{2} \left( \left( \frac{d x}{dt} \right)^2 + \omega(t)^2 x^2 \right),
\eeq
where $\omega(t)>0$ is time-dependent with $\omega(t) \to \omega_\pm$ for $ t \to \pm \infty$.
The operator $x(t)$ can be written by using creation and annihilation operators as
\beq
x(t) \to \left \{ \begin{array}{ll}
\frac{1}{\sqrt{2\omega_-}} ( a e^{- i\omega_- t} +a^\dagger e^{ i\omega_- t}) \quad& t \to -\infty \\
\frac{1}{\sqrt{2\omega_+}} ( b e^{- i\omega_+ t} +b^\dagger e^{ i\omega_+ t}) \quad& t \to +\infty
\end{array}
\right. \label{eq:inftyexp}
\eeq
where $[a,a^\dagger]=1$ and $[b,b^\dagger]=1$ as usual. The point of the time dependence of $\omega(t)$ is that 
these creation/annihilation operators at $t \to \pm \infty$ are related by a nontrivial Bogoliubov transformation as
\beq
a = \alpha b - \beta b^\dagger, \qquad a^\dagger = \alpha^* b^\dagger - \beta^* b.
\eeq
Here $\alpha$ and $\beta$ are constants which are determined by solving the equation of motion $\frac{d^2}{dt^2}x(t) + \omega(t)^2 x(t) =0$
and comparing the limits $t \to \pm \infty$. But the explicit computation is not necessary for our purposes.
For the commutation relations $[a,a^\dagger]=1$ and $[b,b^\dagger]=1$ to be consistent, we have $|\alpha|^2 - |\beta|^2=1$.

The state $\ket{\Omega_{t=-\infty}}$ is defined by
$
a \ket{\Omega_{t=-\infty}} =0.
$
This means $(\alpha b - \beta b^\dagger) \ket{\Omega_{t=-\infty}} =0$ and hence we get
\beq
\ket{\Omega_{t=-\infty}} = C \sum_{k \geq 0}\left( \frac{\beta}{\alpha} \right)^k \sqrt{\frac{(2k)!}{2^{2k}(k!)^2}} \cdot \ket{2k}_{t=+\infty} \label{eq:transf}
\eeq
where 
$
\ket{n}_{t=+\infty}:=  (n!)^{-1/2} (b^\dagger)^{n}\ket{\Omega_{t=+\infty}} 
$, and 
$C$ is an overall normalization constant. 
By Stirling formula we have
\beq
\sqrt{\frac{(2k)!}{2^{2k}(k!)^2}}  \sim  (\pi k)^{-1/2} \qquad (k \gg1).
\eeq
Also we introduce
$
\beta / \alpha= e^{- \eta + i \theta} 
$,
where $\eta>0$ and $\theta$ are real numbers. The $\eta$ is positive because $|\alpha|^2 - |\beta|^2=1$.
Then we get
\beq
\ket{\Omega_{t=-\infty}} \sim C \sum  e^{ (- \eta  + i \theta) k}  (\pi k)^{-1/2} \ket{2k}_{t=+\infty} \
\eeq
for large $k$. Now it is clear that this state satisfies the condition that $\exp( \epsilon H_{t=+\infty} ) \ket{\Omega_{t=-\infty}}$
has a finite norm as long as $\epsilon < \eta/2\omega_+$. Notice that $\eta/\omega_+$ is like an inverse temperature in a rough sense.

In the above discussion we assumed that the Hamiltonian is time-independent in the region $t \to \pm \infty$.
However, what we have seen is essentially the fact that Bogoliubov transformations do not spoil the condition of the Reeh-Schlieder theorem
(at least in the above simple harmonic oscillator). The effects of backgrounds are that, intuitively speaking, we have successive Bogoliubov transformations
as the time evolves.
We expect that this is a general feature. An intuition is that the background fields affect low energy modes but not high energy modes
(as far as the background fields are smooth so that the Fourier transform of the background fields are exponentially suppressed at high energies), and
the condition of the Reeh-Schlieder theorem is only about the high energy modes.

One may think that black holes are fundamentally different because it has the spacetime singularity.
However, just the presence of the singularity may not imply the violation of the Reeh-Schlieder theorem.
To argue this point, let us recall how the Hartle-Hawking state~\cite{Hartle:1976tp} is defined.

The metric of a Schwarzchild black hole is 
\beq
ds^2 = - f(r) dt^2 + \frac{1}{f(r)} dr^2 +r^2 d\Omega^2_{d-2}
\eeq
where $f(r)$ is a function of $r$ which is positive in the region $r>r_s$, negative in $0<r<r_s$, and has a singularity at $r=0$.
The $f(r)$ becomes zero at the horizon $r=r_s$ with $f'(r_s) \neq 0$.
We introduce the tortoise coordinate $r_*$ and the Kruskal coordinates $U$ and $V$ as
\beq
r_* &=  \int \frac{dr}{f(r)}  \\
U &= - \exp \left(  \frac{1}{2} f'(r_s)(- t +r_*) \right) \\
V&  =  \exp \left( \frac{1}{2} f'(r_s)( t +r_*) \right) .
\eeq
By using them, the metric becomes
\beq
ds^2 = - \frac{4f(r)}{f'(r_s)^2\exp(f'(r_s)r_*)} dUdV +r^2 d\Omega^2_{d-2}.
\eeq
The $r$ is determined from $U$ and $V$ as $\exp(f'(r_s)r_*)=-UV$, and one can check that this metric is smooth in the region $r>0$.
This is the metric of a two-sided black hole.

The Hartle-Hawking state $\ket{{\rm HH}}$ is defined as follows. First we analytically continue the above metric 
as $U \to - \overline{Z}$ and $V \to Z$. Here $Z$ and $\overline{Z}$ are two independent complex coordinates, and $r$ is still defined by 
$\exp(f'(r_s)r_*)=Z\overline{Z}$ as an analytic function of $Z$ and $\overline{Z}$. Now, we restrict these coordinates to a real analytic submanifold 
\beq
\CM &=\{(Z,\overline{Z}, \Omega_{d-2});~\overline{Z}=Z^*\} \\
ds^2 &= \frac{4f(r)}{f'(r_s)^2\exp(f'(r_s)r_*)} |dZ|^2 +r^2 d\Omega^2_{d-2}.
\eeq
where $Z^*$ means the complex conjugate of $Z$.
The geometry is completely smooth since the singularity $r=0$ is away from this submanifold. This submanifold has a smooth metric of Euclidean signature.
We perform the path integral on this submanifold, and then analytically continue the results back to the original coordinates $U$ and $V$.

More explicitly, we may compute correlation functions on the Euclidean manifold $\CM$ and then analytically continue the result to the original Schwarzchild background.\footnote{
This analytic continuation is not Wick rotation if we want to get Wightman correlation functions instead of time-ordered correlation functions.
For the Reeh-Schlieder theorem we need Wightman correlation functions. Let us recall how it is done in the case of the flat space.
First we consider a Euclidean correlation function $\vev{\phi(x_1^{\rm E} ) \cdots \phi(x^{\rm E}_N)}$ where the superscript E means ``Euclidean".
Let $\tau_1, \cdots, \tau_N$ be the Euclidean time coordinates of $x^{\rm E}_1, \cdots, x^{\rm E}_N$. We analytically continue them
as $\tau_i \to z_i = \tau_i+i t_i $. Then, we take the time ordering of the Euclidean times $\tau_i$ (but not the Minkowski times $t_i$) as
$\tau_1>\tau_2 > \cdots > \tau_N$. After this ordering, we take $\tau_1 \to 0$, keeping $0>\tau_2> \cdots >\tau_N$. Then we take $\tau_2 \to 0$, and then $\tau_3 \to 0$, and so on.
After taking $\tau_N \to 0$, we get the desired Wightman correlation function. Compare this process to the proof of the Reeh-Schlieder theorem.
We do not try to work out the details in the case of the Hartle-Hawking state, but the idea should be similar.}
In this way, we get correlation functions on the Hartle-Hawking state, 
\beq
\bra{{\rm HH}} \phi(x_1) \cdots \phi(x_N) \ket{{\rm HH}}.
\eeq
From these correlation functions, the state $\ket{{\rm HH}}$ itself and the entire Hilbert space $\CH$ may be recovered
by the Wightman reconstruction theorem~\cite{Streater:1989vi} (assuming its validity on this curved spacetime).

By using the first $K$ operators to create another state as 
\beq
\bra{\chi} = \int d^dx_1 \cdots d^dx_K \bra{{\rm HH}} f_1(x_1)\phi(x_1) \cdots f_K(x_K)\phi(x_K)  
\eeq
we get matrix elements of the form
\beq
\bra{\chi}\phi(x_{K+1}) \cdots \phi(x_N) \ket{{\rm HH}}.\label{eq:HHelement}
\eeq
This is the type of matrix elements which appeared in the proof of the Reeh-Schlieder theorem.

From the beginning, the matrix elements \eqref{eq:HHelement} are obtained from analytically continued correlation functions.
Therefore, it is very likely that the Reeh-Schlieder theorem holds in the Hartle-Hawking state.

In the case that the black hole is in an asymptotically AdS spacetime and the theory has a CFT dual, the Hartle-Hawking state corresponds to 
a thermofield state \cite{Maldacena:2001kr}
\beq
\ket{\Psi} \propto \sum e^{-\beta E_i/2} \ket{i} \otimes \ket{i^*} \label{eq:double}
\eeq
where $\ket{i}$ spans the basis of the CFT Hilbert space $\CH(S^{d-1})$ on $S^{d-1}$,
and $\ket{i^*}$ are the dual basis of the Hilbert space $\CH(\overline{S^{d-1}})$ on the orientation-flipped sphere $\overline{S^{d-1}}$.\footnote{
Although it is not relevant for the following discussion, we would like to make a remark about how to think about them.
A spatial section of the boundary of the AdS-Schwarzchild geometry is $S^{d-1} \sqcup \overline{S^{d-1}}$, where $\sqcup$ means disjoint union of the two manifolds.
Therefore, we get the Hibert space of a single theory on $S^{d-1} \sqcup \overline{S^{d-1}}$ (instead of two copies of the theory on $S^{d-1}$).
By an axiom of quantum field theory, $\CH(Y_1 \sqcup Y_2) \cong \CH(Y_1) \otimes \CH(Y_2)$ for any disjoint $Y_1$ and $Y_2$,
and hence $\CH(S^{d-1} \sqcup \overline{S^{d-1}}) \cong \CH(S^{d-1}) \otimes \CH( \overline{S^{d-1}}) $.
Another axiom of quantum field theory states $ \CH( \overline{Y} ) \cong \overline{\CH(Y)}$, where the overline on the Hilbert space means
the complex conjugate vector space. Thus, given a state $\ket{i} \in \CH(Y)$, we naturally get a corresponding state $\ket{i^*} :=\overline{\ket{i}} \in \CH(\overline{Y})$.
The overline on the spatial manifold $\overline{Y}$ is basically an orientation flip, but it can be more complicated depending on whether one considers
spin theory, pin$^\pm$ theory, and so on. See~\cite{Freed:2016rqq,Yonekura:2018ufj} for more details.
} The $E_i$ is the energy eigenvalue of both $\ket{i}$ and $\ket{i^*}$, and $\beta$ is the inverse temperature of the black hole.
In the CFT, this state satisfies the condition that $e^{\epsilon H} \ket{\Psi}$ has a finite norm
for $\epsilon < \beta/4$.

Therefore, the existence of the singularity itself does not seem to violate the Reeh-Schlieder theorem.
In the Hartle-Hawking state the Reeh-Schlieder theorem is very likely to hold.
Any smooth deformation from the Hartle-Hawking background seems to satisfy the condition of the Reeh-Schlieder theorem,
because of the intuition gained from the harmonic oscillator.

\section{Information recovery in black hole evaporation: A proposal}
\begin{figure}
\centering
\includegraphics[width=0.75\textwidth]{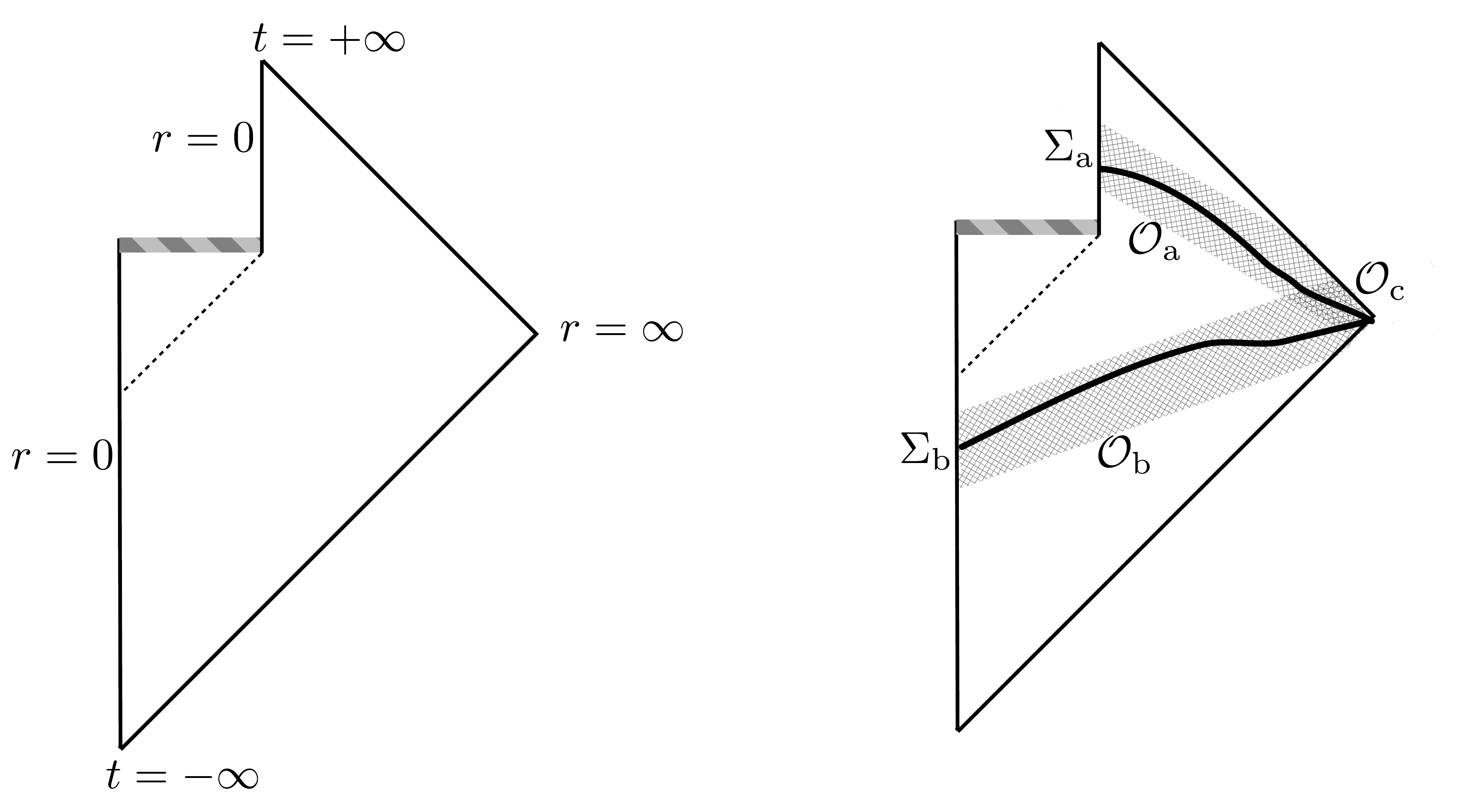}
\caption{Left: A Penrose diagram of an evaporating black hole. The thick gray line is the spacetime singularity, and the dotted line is the event horizon.
We omit to write the matter which cause the gravitational collapse.
Right: Two Cauchy surfaces $\Sigma_{\rm b}$ and $\Sigma_{\rm a}$ which are homotopically inequivalent if we require the Cauchy surface to be away from the singularity. 
We also take their tubular neighborhoods denoted as $\CO_{\rm b}$ and $\CO_{\rm a}$, and their intersection is denoted as $\CO_{\rm c}:=\CO_{\rm a} \cap \CO_{\rm b} $.
The region $\CO_{\rm c}$ may be considered to be a ``neighborhood of spatial infinity".
 \label{fig:evaporate}}
\end{figure}

Now we would like to discuss how a small amount of information which is thrown into a black hole may be recovered by using the Reeh-Schlieder theorem.
We always work in the approximation in which the background geometry is fixed and we just consider perturbation around that background.
This is necessary in order to fix the gauge associated to diffeomorphisms and the concept of local operators to make sense. 
To make it more concrete, we may use BRST quantization of gravity in which the Hilbert space is enlarged to incorporate unphysical modes. In the proof of the Reeh-Schlieder theorem,
the positive definiteness of the inner product in the Hilbert space may not be necessary. The inner product needs to be non-degenerate, which is presumably 
satisfied in the total unphysical Hilbert space of BRST quantization. Admittedly, the concept of algebras $\CA(\CO)$ of
not-necessarily BRST invariant local operators is not so beautiful. However, the final conclusion of this section is that ``information" can be stored near the region 
of spatial infinity $r \to \infty$. If we take the limit $r \to \infty$, hopefully the BRST transformation does not matter there because the gauge transformation
is assumed to be trivial at $r \to \infty$. Alternatively, we may consider BRST invariant but not completely local operators by starting from spatial infinity,
which are analogous to operators of the form $\psi(x) \exp( i \int^x_\infty A)$ in gauge theories. This may be possible if the region $\CO$ contains $r \to \infty$, which is the case
in the following discussion. (See the region $\CO_c$ in the right side of Figure~\ref{fig:evaporate}.)

We assume that there exists a state $\ket{\Psi}$ which is described by the geometry of the left side of Figure~\ref{fig:evaporate},
and we further assume that the Reeh-Schlieder theorem holds in this state.
In more detail, we are assuming the validity of the effective field theory description in the region $\CO_{\rm b}$ and $\CO_{\rm a}$ in Figure~\ref{fig:evaporate} 
where we will apply the Reeh-Schlieder theorem to $\ket{\Psi}$. (See the next paragraph for explanation of $\CO_{\rm b}$ and $\CO_{\rm a}$.)
We emphasize that the effective field theory need not be valid near the singularity. In the following discussion,
we completely avoid the region near the singularity.
The validity of the effective field theory in $\CO_{\rm b} \cup \CO_{\rm a}$ is debetable,
especially because of the severe backreaction of the Hawking radiation, and
the author does not have strong argument for it. It would be very interesting
to prove/disprove this assumption,
but we just assume it in this paper.

More precisely, we need the following assumption. Let us take two Cauchy surfaces $\Sigma_{\rm b}$ and $\Sigma_{\rm a}$
as in the right side of Figure~\ref{fig:evaporate}.  (The subscripts mean ``before" and ``after" the evaporation.)
Next we take tubular neighborhoods of these Cauchy surfaces denoted as 
$\CO_{\rm b}$ and $\CO_{\rm a}$ such that they are far away from the spacetime singularity.
The open set $\CO_{\rm b}$ can be taken to be outside of the event horizon if one wishes so. 
Moreover, we can take them so that the intersection $\CO_{\rm c}:= \CO_{\rm b} \cap \CO_{\rm a}$ is not empty, $\CO_{\rm c} \neq \varnothing$.
This $\CO_{\rm c}$ is located in a region which is space-like separated from the singularity: see Figure~\ref{fig:evaporate}.
Then we assume that
\begin{enumerate}
\item There exists a Heisenberg picture state $\ket{\Psi}$ describing the situation of Figure~\ref{fig:evaporate}.
\item The states of the form $ b\ket{\Psi}$ for $b \in \CA(\CO_{\rm b})$ are dense in the Hilbert space of states before the black hole evaporation.
\item The states of the form $ a\ket{\Psi}$ for $a \in \CA(\CO_{\rm a})$ are dense in the Hilbert space of states after the black hole evaporation.
\end{enumerate}
The second and third assumptions are a version of the ``time-slice axiom"; see the axiom G. ``Time-slice Axiom" of Section II.1.2 of \cite{Haag:1992hx}.

To understand what we are assuming here, it is helpful to consider the case of free field theories.
A free field theory can be quantized by the standard canonical quantization in the region $\CO_{\rm b}$, and we can construct the Hilbert space $\CH_{\rm b}$ as a Fock space.
We can also quantize the free field theory in the region $\CO_{\rm a}$ and obtain the Hilbert space $\CH_{\rm a}$. These are the Hilbert spaces before and after the
black hole evaporation. Then we assume the existence of state vectors $\ket{\Psi}_{\rm b} \in \CH_{\rm b}$ and $\ket{\Psi}_{\rm a} \in \CH_{\rm a}$ which physically describe a single state $\ket{\Psi}$
in the respective Hilbert spaces. For example, if the free field theory is the $1+0$-dimensional harmonic oscillator discussed in Sec.~\ref{sec:Curved},
the quantization is performed by using \eqref{eq:inftyexp}, and an example of the state $\ket{\Psi}$ is given by 
\beq
\ket{\Psi}_{\rm b} &=\ket{\Omega_{t=-\infty}}, \\
\ket{\Psi}_{\rm a} & = C \sum_{k \geq 0}\left( \frac{\beta}{\alpha} \right)^k \sqrt{\frac{(2k)!}{2^{2k}(k!)^2}} \cdot  ((2k)!)^{-1/2} (b^\dagger)^{2k}\ket{\Omega_{t=+\infty}} .
\eeq
See \eqref{eq:transf}. In the following, we will identify $\CH_{\rm b}$ and $\CH_{\rm a}$ by using the Reeh-Schlieder theorem.

How the state $\ket{\Psi}$ ``looks like" before the evaporation of the black hole (i.e., $\ket{\Psi}_{\rm b}$) is just determined by the initial state of the gravitational collapse. For example, 
we may take it to be a collapsing big star.

On the other hand, how the state $\ket{\Psi}$ ``looks like" after the evaporation (i.e., $\ket{\Psi}_{\rm a}$) is a highly nontrivial version of the problem which is
analogous to the one considered in the harmonic oscillator 
as in \eqref{eq:transf}. Basically it consists of the Hawking radiation and some Planck scale effects, and its computation requires a full theory of quantum gravity which can treat the singularity.
However we do not need its explicit form in the following discussion. We just assume its existence.

Now we consider another state $\ket{\Phi}_{\rm b} \in \CH_{\rm b}$ in which we throw additional matter to the black hole described by $\ket{\Psi}_{\rm b}$.
We assume that the amount of the matter is small so that the backreaction to the geometry is small and treated in perturbation theory (in BRST quantization).
By the time-slice axiom, this state is approximated by a state of the form  $ b\ket{\Psi}_{\rm b}$ for $b \in \CA(\CO_{\rm b})$ to an arbitrary good accuracy. 

If we look back the proof of the Reeh-Schlieder theorem, what was shown there is that
the space $\{  c\ket{\Psi}_{\rm b};\  c \in \CA(\CO_{\rm c}) \}$ is dense in the space $\{  b\ket{\Psi}_{\rm b};\ b \in \CA(\CO_{\rm b}) \}$.
By the time-slice axiom, this space $\{  b\ket{\Psi}_{\rm b};\ b \in \CA(\CO_{\rm b}) \}$ is dense in the Hilbert space $\CH_{\rm b}$
(at least in the approximation of the fixed background geometry).
Thus our state $\ket{\Phi}_{\rm b}$ is very well described by a state of the form $c \ket{\Psi}_{\rm b}$ to an arbitrary good accuracy.
Now, because $\CO_{\rm c} \subset \CO_{\rm a}$, we have $c \in \CA(\CO_{\rm c}) \subset \CA(\CO_{\rm a}) $.
Then we notice that the state $c \ket{\Psi}_{\rm a}$ makes sense as a state vector in the Hilbert space $\CH_{\rm a}$ after the evaporation of the black hole.
We propose that this is indeed the state vector after the black hole evaporation.
Therefore, we get the pure state $c \ket{\Psi}_{\rm a}$ without any loss of the information of the additional matter thrown into the black hole,
once we assume that $\ket{\Psi}_{\rm a}$ itself is pure. As discussed in the introduction, a big black hole may be reached by successively throwing a small amount of matter to a smaller black hole,
and hence we may reduce the information problem to a Planck size black hole.

In more detail, we obtain the linear map
\beq
\CH_{\rm b} \ni c\ket{\Psi}_{\rm b} \mapsto c\ket{\Psi}_{\rm a} \in \CH_{\rm a}. \label{eq:Umap}
\eeq
We assume that this also preserves the inner product. The Reeh-Schlieder theorem implies that any state in a dense subset of $\CH_{\rm b}$ can be uniquely written as $c\ket{\Psi}_{\rm b}$ (i.e.
the state $\ket{\Psi}_b$ is cyclic and separating with respect to $\CA(\CO_{\rm c})$), and hence the above map is well-defined and can be completed to a unitary map $\CH_{\rm a} \to \CH_{\rm b}$.
Under this unitary map, we identify $\CH_{\rm b}$ and $\CH_{\rm a}$ and get a single Hilbert space $\CH$. This $\CH$ is the Heisenberg picture Hilbert space.

Notice that the Reeh-Schlieder theorem is just an existence proof of the operator $c \in \CA(\CO_{\rm c})$ and it does not tell us how the operator $c$ looks like.
The fact that the region $\CO_{\rm c}$ is far away from the black hole although we are thinking about throwing something into the black hole
means that this operator can be extremely complicated. 
One may call it a version of scrambling of the information.  In any case, via the operator $c \in \CA(\CO_{\rm c})$
we may connect the world before and after the black hole evaporation as in \eqref{eq:Umap}, without requiring any radical new ideas.
The Reeh-Schlieder theorem is already radical enough.

The essential point of the above discussion is analyticity. Roughly speaking,
in a state with finite energy, quantum fields are analytic.
We have analytic fields in the region $\CO_{\rm b}$, which is analytically continued to the region $\CO_{\rm a}$ through the region $\CO_{\rm c}$.
By this analytic continuation, we get a unique state after the evaporation.
See Appendix~\ref{app:wave} for a simple demonstration of the analyticity in quantum mechanics.

We also emphasize that we have only used the regions $\CO_{\rm b} \cup \CO_{\rm a}$ in the spacetime.
What happens inside the event horizon and at the singularity is irrelevant, except that 
we have assumed the existence of $\ket{\Psi}_{\rm a}$ for a specific single state $\ket{\Psi}$.
Then, for any other initial state $\ket{\Phi}_{\rm b}$, the state $\ket{\Phi}_a$ after evaporation can be constructed by the above argument .

If we apply the above arguments to approximate global symmetry charges such as the baryon number of the standard model,
one gets an apparent contradiction whose resolution may suggest some properties of the black hole evaporation.
We leave the discussion of this point to future work.

Finally, let us notice an important analogy with the AdS/CFT correspondence.
The AdS/CFT correspondence says that we can generate all states in the Hilbert space by acting boundary CFT operators to a state (in which the Reeh-Schlieder theorem holds).
Indeed, the region $\CO_{\rm c}$ is near the spatial infinity $r \to \infty$ as is clear from Figure~\ref{fig:evaporate}.
The analogy becomes even more clear if we put a black hole in AdS and consider a version of the AdS/CFT dictionary between bulk operators $\phi(r,x)$ and boundary operators $\phi(x)$
given by $\lim_{r \to \infty} r^\Delta \phi(r,x) = \phi(x)$~\cite{Banks:1998dd,Harlow:2011ke}.
Therefore, our discussion is essentially the same as the argument that there is no information loss in CFT and hence no information loss in AdS.
We can create all (or more precisely dense) states by operators near spatial infinity. 
The point of our discussion is that we are able to discuss the absence of information loss without using the CFT dual,
but instead by using the Reeh-Schlieder theorem and the time-slice axiom in the bulk. Indeed, our discussion also makes sense (in the approximation of a fixed background) in asymptotically flat spacetimes in which
the existence of a dual theory is not clear. (See e.g. \cite{Nomura:2016ikr} for an attempt towards this direction.)

By using the above understanding, the ER=EPR proposal of \cite{Maldacena:2013xja} has a natural justification,
where ER means Einstein-Rosen bridge and EPR means Einstein-Podolsky-Rosen entanglement. 
Let us discuss it in the most basic case of a two-sided AdS-Schwartzchild black hole.
We start from the boundary of one side of the black hole. Consider boundary operators on that boundary. We slightly move them into the bulk by the dictionary mentioned above.
Then, the bulk Reeh-Schlieder theorem says that we can generate a dense subspace of the Hilbert space by these operators. We can move the region of the support of these operators
to the other side by going through the worm hole. Then we reach the other boundary. This consideration implies that the state $\ket{\Psi}$ describing the worm hole
has the following property. For an arbitrary given operator $a'$ on one of the boundaries, we can find an operator $a''$ on the other boundary
such that $a'\ket{\Psi}$ is well approximated by $a''\ket{\Psi}$ to an arbitrary good accuracy. This is possible only if the two boundaries are fully entangled with each other.
Here it is crucial that the two sides are connected (by the worm hole), since the proof of the Reeh-Schlieder theorem requires analytic continuation from a local region to a tubular neighborhood of a Cauchy surface.
In summary, we can say that when the geometry is connected, the two boundaries are entangled by the Reeh-Schlieder theorem. In this basic case
the state is explicitly given by \eqref{eq:double} which is clearly entangled, but the above discussion is more general (though less quantitative).
This is our interpretation of ER=EPR.

\acknowledgments
I would like to thank, Y.~Tachikawa and Y.~Tanimoto for helpful criticisms and discussions.
The work of K.Y.~is supported by JSPS KAKENHI Grant-in-Aid (Wakate-B), No.17K14265.

\appendix

\section{Finite energy, analyticity, and non-localizability of wave functions}\label{app:wave}
Some of the essential points of the Reeh-Schlieder theorem can already be seen in quantum mechanics in an elementary way.
For the purpose of a simple demonstration, let us consider a wave function $\psi(x)$ where $x \in \BR$ is a space coordinate.
The Fourier transform of it is denoted as
\beq
\widetilde{\psi}(k) = \int dx e^{- i k x} \psi(x).
\eeq
Now, we say that the state described by $\psi(x)$ has ``finite energy" (or more precisely finite momentum in this case) 
if the following technical condition is satisfied:
\beq
\exists \epsilon >0 \text{ such that } \lim_{|k| \to \infty} |\widetilde{\psi}(k)| e^{\epsilon |k|} = 0.
\eeq
This condition means that the wave function $\widetilde{\psi}(k)$ in momentum space decays sufficiently quickly at large momentum regions.\footnote{
If the potential $V(x)$ of the quantum mechanical system is not smooth, a state with $H \psi = E \psi$ for finite $E$ does not necessarily satisfy
the ``finite energy" condition in the above sense. So we assume that $V(x)$ is analytic.
In the context of this paper, $V(x)$ may be regarded as gravitational potential. If the total system including dynamical gravity has finite energy,
gravitational fields and hence $V(x)$ should be analytic.
}

If the finite energy condition in the above sense is satisfied, we can show that the position space wave function $\psi(x)$ is an analytic function.
To show it, we write it as
\beq
\psi(x) &= \int \frac{dk}{2\pi} e^{ i k x} \widetilde{\psi}(k) \nonumber \\
&= \int \frac{dk}{2\pi} e^{ i k x - \epsilon |k|} [ e^{\epsilon |k|} \widetilde{\psi}(k) ].
\eeq
The product $e^{\epsilon |k|} \widetilde{\psi}(k)$ is bounded as $|e^{\epsilon |k|} \widetilde{\psi}(k) | <C$ for some $C$.
Thus the above integral is absolutely convergent as long as the imaginary part of $x$ is in the region
\beq
 - \epsilon < \Im (x) < \epsilon. \label{eq:complexR}
\eeq
Therefore, we can analytically continue the wave function $\psi(x)$ to the complex region given by \eqref{eq:complexR}.
In particular, $\psi(x)$ is analytic on the real axis $x \in \BR$. This completes the proof.

The states satisfying the finite energy condition are dense in the Hilbert space. Moreover, reasonable physical systems directly satisfy the condition.
For example, for a relativistic particle with energy $E_k \sim \sqrt{m^2+k^2}$, any state which looks like a thermal state $e^{-\beta H}$ satisfies the finite energy condition
by taking $\epsilon $ as $\epsilon <\beta/2$. This is because the finite energy condition can be formulated as $\int dk e^{-2\epsilon |k|}  |\widetilde{\psi}(k)|^2 < \infty $,
which (by replacing $\psi \psi^\dagger$ by $e^{ - \beta H}$) is given as $\int dk e^{-2\epsilon |k|}  e^{- \beta E_k} < \infty $.
For a system to violate the finite energy condition, the temperature $T=\beta^{-1}$ must be infinity, which is quite unphysical. 

\begin{figure}
\centering
\includegraphics[width=0.9\textwidth]{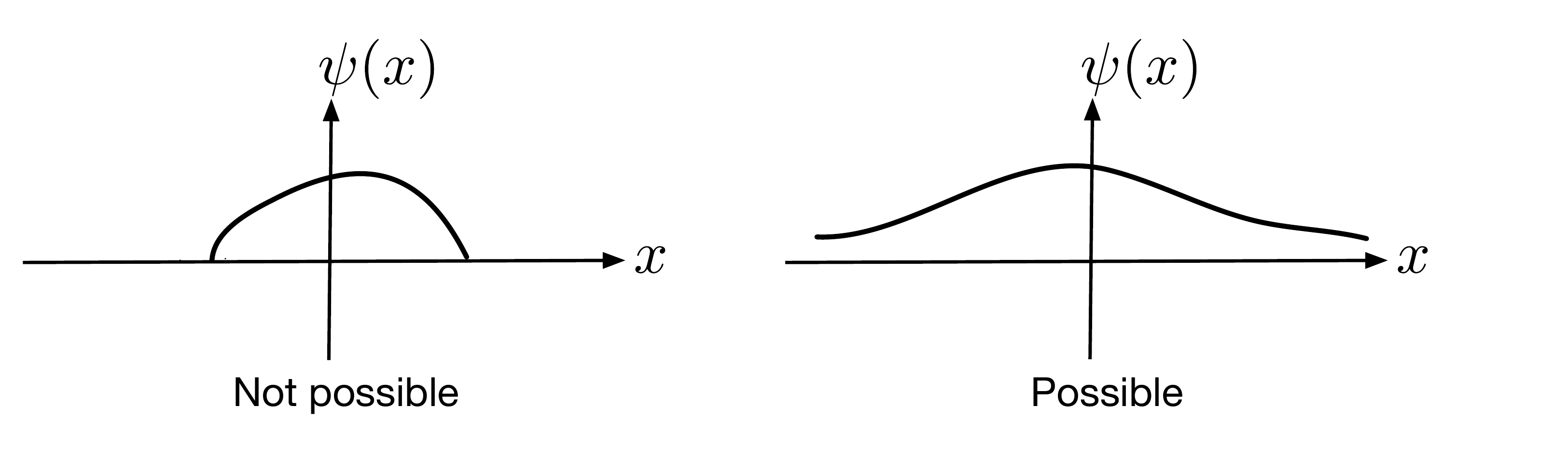}
\caption{ For an analytic wave function, a completely localized wave function like the left figure is not possible.
What is possible is like the right figure. The tail of the wave function may be extremely small. Nevertheless,
the entire wave function is determined from the tail by analytic continuation.
 \label{fig:loc}}
\end{figure}

The implication of the theorem is that the wave function is not completely localized.
See Figure~\ref{fig:loc}. An analytic wave function always has a tail.
No matter how small the tail is, the entire wave function is determined by analytic continuation from the tail. 
In the context of the present paper, we may roughly say that ``the wave function inside a black hole is determined by analytic continuation from the wave function
outside it." It is completely a different question whether we can determine the wave function by measurement or not. Probably we cannot.
But the problem discussed in this paper is not about measurement, but about the time evolution of a evaporating black hole.
For that purpose, the above fact really implies that information is not localized.

Finally, let us mention the relation between the Reeh-Schlieder theorem and the theorem discussed in this appendix.
In the proof of the Reeh-Schlieder theorem sketched in Sec.~\ref{sec:RS}, it is crucial that matrix elements of the form
\beq
\bra{\Phi} \phi(x_1) \cdots \phi(x_n) \ket{\Psi}
\eeq
are analytic if $\ket{\Psi}$ satisfy the finite energy condition $\bra{\Psi} e^{2\epsilon H} \ket{\Psi} < \infty$. 
In fact, if we take $\ket{\Phi}$ to be a ground state $\ket{\Omega}$, we may interpret it as a ``many body wave function"
\beq
\psi(x_1, \cdots, x_n) \sim \bra{ \Omega } \phi(x_1) \cdots \phi(x_n) \ket{\Psi}.
\eeq
Thus the main point of both of the theorems in this appendix and the Reeh-Schlieder theorem is that 
finite energy implies analyticity of wave functions and quantum fields.


\bibliographystyle{JHEP}
\bibliography{ref}

\providecommand{\href}[2]{#2}\begingroup\raggedright\begin{thebibliography}{10}

\bibitem{Hawking:1976ra}
S.~W. Hawking, {\it {Breakdown of Predictability in Gravitational Collapse}},
  {\em Phys. Rev.} {\bf D14} (1976) 2460--2473.

\bibitem{Hawking:1974sw}
S.~W. Hawking, {\it {Particle Creation by Black Holes}},  {\em Commun. Math.
  Phys.} {\bf 43} (1975) 199--220. [,167(1975)].

\bibitem{RS}
H.~Reeh and S.~Schlieder, {\it {Bemerkungen zur Unita\"{a}r\"{a}quivalenz von
  Lorentzinvarienten Feldern}},  {\em Nuovo Cimento} {\bf {\bf 22}} (1961)
  1051.

\bibitem{Streater:1989vi}
R.~F. Streater and A.~S. Wightman, {\em {PCT, spin and statistics, and all
  that}}.
\newblock 1989.

\bibitem{Witten:2018zxz}
E.~Witten, {\it {Notes on Some Entanglement Properties of Quantum Field
  Theory}},  \href{http://arxiv.org/abs/1803.04993}{{\tt arXiv:1803.04993}}.

\bibitem{Papadodimas:2013wnh}
K.~Papadodimas and S.~Raju, {\it {Black Hole Interior in the Holographic
  Correspondence and the Information Paradox}},  {\em Phys. Rev. Lett.} {\bf
  112} (2014), no.~5 051301, [\href{http://arxiv.org/abs/1310.6334}{{\tt
  arXiv:1310.6334}}].

\bibitem{Papadodimas:2013jku}
K.~Papadodimas and S.~Raju, {\it {State-Dependent Bulk-Boundary Maps and Black
  Hole Complementarity}},  {\em Phys. Rev.} {\bf D89} (2014), no.~8 086010,
  [\href{http://arxiv.org/abs/1310.6335}{{\tt arXiv:1310.6335}}].

\bibitem{Verlinde:2013qya}
E.~Verlinde and H.~Verlinde, {\it {Behind the Horizon in AdS/CFT}},
  \href{http://arxiv.org/abs/1311.1137}{{\tt arXiv:1311.1137}}.

\bibitem{Strohmaier:2002mm}
A.~Strohmaier, R.~Verch, and M.~Wollenberg, {\it {Microlocal analysis of
  quantum fields on curved space-times: Analytic wavefront sets and
  Reeh-Schlieder theorems}},  {\em J. Math. Phys.} {\bf 43} (2002) 5514--5530,
  [\href{http://arxiv.org/abs/math-ph/0202003}{{\tt math-ph/0202003}}].

\bibitem{Gerard:2017apb}
C.~G\'erard and M.~Wrochna, {\it {Analytic Hadamard states, Calder\'on
  projectors and Wick rotation near analytic Cauchy surfaces}},
  \href{http://arxiv.org/abs/1706.08942}{{\tt arXiv:1706.08942}}.

\bibitem{Sanders:2008gs}
K.~Sanders, {\it {On the Reeh-Schlieder Property in Curved Spacetime}},  {\em
  Commun. Math. Phys.} {\bf 288} (2009) 271--285,
  [\href{http://arxiv.org/abs/0801.4676}{{\tt arXiv:0801.4676}}].

\bibitem{Morrison:2014jha}
I.~A. Morrison, {\it {Boundary-to-bulk maps for AdS causal wedges and the
  Reeh-Schlieder property in holography}},  {\em JHEP} {\bf 05} (2014) 053,
  [\href{http://arxiv.org/abs/1403.3426}{{\tt arXiv:1403.3426}}].

\bibitem{Haag:1992hx}
R.~Haag, {\em {Local quantum physics: Fields, particles, algebras}}.
\newblock 1992.

\bibitem{Harlow:2014yka}
D.~Harlow, {\it {Jerusalem Lectures on Black Holes and Quantum Information}},
  {\em Rev. Mod. Phys.} {\bf 88} (2016) 015002,
  [\href{http://arxiv.org/abs/1409.1231}{{\tt arXiv:1409.1231}}].

\bibitem{Almheiri:2012rt}
A.~Almheiri, D.~Marolf, J.~Polchinski, and J.~Sully, {\it {Black Holes:
  Complementarity or Firewalls?}},  {\em JHEP} {\bf 02} (2013) 062,
  [\href{http://arxiv.org/abs/1207.3123}{{\tt arXiv:1207.3123}}].

\bibitem{Papadodimas:2012aq}
K.~Papadodimas and S.~Raju, {\it {An Infalling Observer in AdS/CFT}},  {\em
  JHEP} {\bf 10} (2013) 212, [\href{http://arxiv.org/abs/1211.6767}{{\tt
  arXiv:1211.6767}}].

\bibitem{Maldacena:2013xja}
J.~Maldacena and L.~Susskind, {\it {Cool horizons for entangled black holes}},
  {\em Fortsch. Phys.} {\bf 61} (2013) 781--811,
  [\href{http://arxiv.org/abs/1306.0533}{{\tt arXiv:1306.0533}}].

\bibitem{Bousso:2012as}
R.~Bousso, {\it {Complementarity Is Not Enough}},  {\em Phys. Rev.} {\bf D87}
  (2013), no.~12 124023, [\href{http://arxiv.org/abs/1207.5192}{{\tt
  arXiv:1207.5192}}].

\bibitem{Harlow:2013tf}
D.~Harlow and P.~Hayden, {\it {Quantum Computation vs. Firewalls}},  {\em JHEP}
  {\bf 06} (2013) 085, [\href{http://arxiv.org/abs/1301.4504}{{\tt
  arXiv:1301.4504}}].

\bibitem{Chakraborty:2017pmn}
S.~Chakraborty and K.~Lochan, {\it {Black Holes: Eliminating Information or
  Illuminating New Physics?}},  {\em Universe} {\bf 3} (2017), no.~3 55,
  [\href{http://arxiv.org/abs/1702.07487}{{\tt arXiv:1702.07487}}].

\bibitem{Hartle:1976tp}
J.~B. Hartle and S.~W. Hawking, {\it {Path Integral Derivation of Black Hole
  Radiance}},  {\em Phys. Rev.} {\bf D13} (1976) 2188--2203.

\bibitem{Maldacena:2001kr}
J.~M. Maldacena, {\it {Eternal black holes in anti-de Sitter}},  {\em JHEP}
  {\bf 04} (2003) 021, [\href{http://arxiv.org/abs/hep-th/0106112}{{\tt
  hep-th/0106112}}].

\bibitem{Freed:2016rqq}
D.~S. Freed and M.~J. Hopkins, {\it {Reflection positivity and invertible
  topological phases}},  \href{http://arxiv.org/abs/1604.06527}{{\tt
  arXiv:1604.06527}}.

\bibitem{Yonekura:2018ufj}
K.~Yonekura, {\it {On the cobordism classification of symmetry protected
  topological phases}},  \href{http://arxiv.org/abs/1803.10796}{{\tt
  arXiv:1803.10796}}.

\bibitem{Banks:1998dd}
T.~Banks, M.~R. Douglas, G.~T. Horowitz, and E.~J. Martinec, {\it {AdS dynamics
  from conformal field theory}},
  \href{http://arxiv.org/abs/hep-th/9808016}{{\tt hep-th/9808016}}.

\bibitem{Harlow:2011ke}
D.~Harlow and D.~Stanford, {\it {Operator Dictionaries and Wave Functions in
  AdS/CFT and dS/CFT}},  \href{http://arxiv.org/abs/1104.2621}{{\tt
  arXiv:1104.2621}}.

\bibitem{Nomura:2016ikr}
Y.~Nomura, N.~Salzetta, F.~Sanches, and S.~J. Weinberg, {\it {Toward a
  Holographic Theory for General Spacetimes}},  {\em Phys. Rev.} {\bf D95}
  (2017), no.~8 086002, [\href{http://arxiv.org/abs/1611.02702}{{\tt
  arXiv:1611.02702}}].

\end{thebibliography}\endgroup

\end{document}